\documentstyle[spconf,amsmath,graphicx,amssymb,xcolor,setspace]{article}
\ninept

\title{FrAUG: A Frame Rate Based Data Augmentation Method \\ for Depression Detection from Speech Signals}
\name{Vijay Ravi $^{1*}$, Jinhan Wang $^1$, Jonathan Flint $^2$, Abeer Alwan $^1$ \thanks{$^*$Corresponding author: vijaysumaravi@ucla.edu}}
\address{$^1$ Dept. of Electrical and Computer Engineering, University of California, Los Angeles, USA \\  $^2$ Dept. of Psychiatry and Biobehavioral Sciences, University of California, Los Angeles, USA}
\begin{document}
\maketitle
\begin{abstract}
In this paper, a data augmentation method is proposed for depression detection from speech signals. Samples for data augmentation were created by changing the frame-width and the frame-shift parameters during the feature extraction process. Unlike other data augmentation methods (such as VTLP, pitch perturbation, or speed perturbation), the proposed method does not explicitly change acoustic parameters but rather the time-frequency resolution of frame-level features. The proposed method was evaluated using two different datasets, models, and input acoustic features. For the DAIC-WOZ (English) dataset when using the DepAudioNet model and mel-Spectrograms as input, the proposed method resulted in an improvement of 5.97\% (validation) and 25.13\% (test) when compared to the baseline. The improvements for the CONVERGE (Mandarin) dataset when using the x-vector embeddings with CNN as the backend and MFCCs as input features were 9.32\% (validation) and 12.99\% (test). Baseline systems do not incorporate any data augmentation. Further, the proposed method outperformed commonly used data-augmentation methods such as noise augmentation, VTLP, Speed, and Pitch Perturbation. All improvements were statistically significant.
\end{abstract}
\section{Introduction}
\label{sec:intro}

Major depressive disorder (MDD), is a common and serious medical illness that negatively affects how one feels, thinks and acts. At its worst, depression can lead to suicide and death. Globally, more than 264 million people are affected by MDD~\cite{james2018global} and by 2030, it is projected to be the second leading cause of disability~\cite{mathers2006projections}. However, only a small percent of these cases get diagnosed and an even smaller percent of them gets treated~\cite{wells1989detection}. Automatic systems for MDD assessment can help reduce diagnostic inequality and will allow for early diagnoses. A possible source of information for building such objective screening mechanisms is the human voice. Among others, the speech signal is an important bio-marker of our mental state~\cite{cummins2015review,ravi2019voice}  and can be collected remotely, in a non-invasive manner with no expert supervision. 

Recently, speech-based automatic diagnosis of depression has gained significant momentum~\cite{alghowinem2013detecting,afshan2018effectiveness,dubagunta2019learning} and advancements in deep learning have pushed their performance  to newer heights~\cite{ma2016depaudionet,he2018automated,harati2021speech,yang2020feature,padi2020multi, rejaibi2019mfcc}. However, data scarcity still remains one of the major challenges in building reliable systems for MDD modeling purposes. Given the sensitivity of mental healthcare data, collection of data can be expensive and challenging. Therefore, there is a need to adopt data augmentation strategies to increase the amount of training data. However, conventional data augmentation techniques (e.g., Vocal-Tract Length Perturbation - VTLP -~\cite{jaitly2013vocal}, Speed and Pitch Perturbation~\cite{ko2015audio}) can be counter-productive when applied to para-linguistic applications such as depression detection because they distort the acoustic data and can lose useful information related to the underlying health condition. 

Previously, Generative Adversarial Network (GAN)~\cite{goodfellow2014generative} based data augmentation was proposed for depression detection~\cite{yang2020feature}. However, GANs themselves require significant amount of training data to be effective.  In~\cite{he2018automated}, spectrograms were rotated to generate new samples and in~\cite{rejaibi2019mfcc}, noise, pitch-shifting and speed perturbation was employed to augment training data. In~\cite{padi2020multi}, a multi-window data augmentation was proposed for emotion recognition which used multiple frame-widths. However, the methods proposed in~\cite{he2018automated, padi2020multi, rejaibi2019mfcc} were not compared with conventional data augmentation techniques and were only evaluated using one model. 

In contrast, in this paper, a frame rate based data augmentation technique is proposed specifically for the task of depression detection from speech signals.  New feature samples were created by varying the frame-width as well as the frame-shift during feature extraction. By changing the frame-rate parameters, the model was provided with different sets of time-frequency resolutions during the training stage. This ensured that acoustic parameters which are thought to correlate with the mental state of the speaker (eg: pitch, formant frequencies, speaking-rate etc.~\cite{cummins2015review,afshan2018effectiveness}) were not inadvertently modified. Additionally, the proposed method was shown to outperform two commonly used data augmentation methods and was validated on two different datasets, input acoustic features and models.

The remainder of this paper is organized as follows: the proposed multi-frame-rate data augmentation approach is introduced in Section~\ref{sec:data_aug}. Experimental details of the datasets and models used are described in Section~\ref{sec:exp_deets}. Results  are reported and discussed in Section~\ref{sec:results_disc} and the conclusion and future directions are provided in Section~\ref{sec:conclusion}.


\section{Proposed Method}
\label{sec:data_aug}



In this section, we describe the proposed data augmentation technique, FrAUG. Given an input speech signal $x[n]$, the windowing and feature extraction process for spectral features, such as spectrograms and mel-frequency cepstral coefficients (MFCC), can be represented as:

\begin{equation}\label{eq:stft}
    X_r[k] = \sum_{m=0}^{L-1} x[m] w[rR-m] e^{-j(2\pi k /N)m} ,
\end{equation}

\noindent where,  $w[rR-m]$  is the sliding window, $r\in \mathbb{Z}$, $N$ is the DFT size, $L$ is the frame-width  and $R$ is the frame-shift~\cite{rabiner2010theory}. The windows overlap by $O = L-R$. $R$ and $O$ are usually specified as a fraction of $L$, which is specified in time or number of samples.


Changing the values of $L$ and $R$ and thereby the frame rate, changes the time-frequency resolution of the extracted features. Conventionally, to balance resolutions between time and frequency, the parameters $L$, $R$ and $O$ are fixed. A Hamming window with $L=25ms$ and $R=40\%$ (i.e $R = 10ms$) is commonly used~\cite{Povey_ASRU2011}.

In FrAUG, given a baseline frame-rate with parameters $(L_1,R_1)$, we augment the training data with features extracted using multiple frame rates with parameters $L_i$ and $R_j$ where $i,j\in \mathbb{N}$. For example, to perform an 8-fold augmentation, frame-widths of $L_2$, $L_3$ and frame-shifts $R_2$, $R_3$ are used along with baseline parameters, resulting in 9 different combinations such as $(L_1,R_2)$,$(L_2,R_3)$,$(L_3,R_2)$, etc. Thus, the model is provided with 8 additional time-frequency resolutions in the training stage. The main advantage of the proposed method is that it does not alter vocal tract or voice source parameters, particularly those useful for depression detection, and is independent of the dataset and model used. Since the underlying mechanism of windowing is the same for non-spectral features (such as the prosodic features), FrAUG can be extended to other acoustic features as well.

\section{Datasets and Models}
\label{sec:exp_deets}

The proposed method was applied on two different models using two distinct datasets and two different input acoustic features. The two models are - DepAudioNet~\cite{ma2016depaudionet} which was trained using mel-spectrograms, and a pretrained x-vector embedding generator trained on MFCCs followed by a convolutional neural network (CNN) backend, for the DAIC-WOZ (English,~\cite{valstar2016avec}) and the CONVERGE (Mandarin,~\cite{li2012patterns}) datasets, respectively. The datasets and the corresponding models used in this paper are described in this section.

\subsection{Datasets}
\label{ssec:datasets}

\subsubsection{DAIC-WOZ}
The Distress analysis interview corpus wizard of Oz (DAIC-WOZ)~\cite{valstar2016avec} database comprises audio-visual interviews of 189 participants, male and female, who underwent evaluation of psychological distress. Each participant was assigned a self-assessed depression score through the patient health questionnaire (PHQ-8) method~\cite{kroenke2009phq}. Audio data belonging only to the participants were extracted using the time-labels provided with the dataset. Recordings from session numbers 318, 321, 341 and 362 were excluded from the training set because of time-labelling errors. This dataset consists of a total of 58 hours of audio data. The data partitioning between train, validation and test sets is the same as that provided with the database description, which is 60\%, 20\% and 20\%, respectively.

\subsubsection{CONVERGE}
The second depression database used in this paper is in Mandarin and was developed as a part of the China, Oxford and Virginia Commonwealth University Experimental Research on Genetic Epidemiology (CONVERGE) study~\cite{li2012patterns}. The CONVERGE study focused on subjects with increased genetic risk for MDD, and to obtain a more genetically homogeneous sample, only women participants were recruited. Each participant was interviewed by a trained interviewer. The diagnoses of depressive (dysthymia and MDD) disorders were made with the Composite International Diagnostic Interview (Chinese version)~\cite{TerSmitten1998}, which classifies diagnoses according to the Diagnostic and Statistical Manual of Mental Disorders fourth edition (DSM-IV) criteria. The database includes recordings of the interviews from 3742 individuals classified as suffering from MDD and 4217 healthy individuals. All audio recordings were collected with a sampling rate of 16kHz. The database was randomly split into 60\%, 20\%, and 20\% for the train, validation, and test sets, respectively. This database contains a total of 391 hours of audio data and is characterized by a large degree of phonetic and content variability. 

\subsection{Models}
\label{ssec:models}

\subsubsection{DepAudioNet}

DepAudioNet is a deep neural network model for detecting depression from speech~\cite{ma2016depaudionet}. This model consisted of a one-dimensional convolutional layer (\textit{Conv1D}) and two unidirectional long short-term memory (LSTM) layers~\cite{bailey2020raw}. The \textit{Conv1D} layer had 40-dimensional mel-spectrograms as input and a kernel size of 3 with no time-dilation. The \textit{Conv1D} was followed by ReLU non-linearity, a dropout layer with $p_{dropout}=0.05$ and a max-pooling layer with kernel size 3. This was followed by two 128-dimensional LSTM layers. Finally, a fully connected layer with Sigmoid activation was applied to generate the binary prediction. 

To mitigate class imbalance between depression and non-depression classes, training data were pre-processed by random cropping and random sampling~\cite{ma2016depaudionet}. First, each utterance was randomly cropped to fragments which were equal to the length of the shortest utterance in the DAIC-WOZ database. This was done in order to negate any bias towards longer durations of audio samples. Each randomly cropped fragment was segmented into 120 frames, where each frame varied between $32ms$ to $128ms$ depending on the frame rate being applied. A training subset was generated by randomly sampling,  without replacement, an equal number of depression and non-depression segments. 

Each experiment consisted of training five separate models for 100 epochs each using a randomly generated training subset. The final prediction was obtained by averaging the probabilities predicted by the five models. The pre-processing step was applied irrespective of the type of data augmentation used and hyperparameters such as batch size, learning rate and the learning rate reduction factor were tuned for every experiment individually. 

This model was used with the DAIC-WOZ dataset. For the baseline experiments, mel-spectrograms were extracted with frame rate parameters of $L=64ms$, and $R=50\%$, as proposed in~\cite{ma2016depaudionet}. When FrAUG was applied, training data was augmented with up to 8 folds using additional frame rates with parameters $L=32ms$ and $L=128ms$ and an overlap $R$ of $25\%$ and $10\%$. The augmentation frame rates were chosen empirically. Even when augmentation was applied, mel-spectrograms for test and validation sets were extracted at the baseline frame rate. 

\subsubsection{x-vector embedding with CNN Classifier}
An x-vector embedding extractor~\cite{snyder2018x} was pre-trained using CN-Celeb~\cite{fan2020cn}, a Mandarin speaker ID dataset. A Kaldi recipe was followed for training the x-vector model~\cite{li2020cn}. Inputs to the x-vector model were 30-dimensional MFCCs and the model consisted of 5 layers of time-dilated convolutional networks followed by average pooling and two fully connected layers; the size of the x-vector embedding was 512. No additional pre-processing was applied. The readers are referred to~\cite{kumar2020designing} for implemetation details. 

After pre-training the x-vector model, embeddings for the CONVERGE dataset were generated which were then used to train a downstream network for classifying depression. The downstream network was made up of two CNN layers followed by two fully connected layers. The downstream model was trained for 100 epochs with a fixed learning rate of $1\mathrm{e}{-4}$. Data augmentation was only applied during the training of the downstream network i.e. embeddings were extracted for the augmented depression training data along with the unaugmented validation and test data. 

Similar to the previous experiment, x-vector embeddings for the baseline experiments were generated using MFCCs extracted with frame rate parameters of $L=64ms$ and $R=50\%$.  When FrAUG was applied, x-vectors were generated using MFCCs extracted with additional frame rate parameters of $L=32ms$, $L=128ms$ and $R=50\%$, $R=25\%$. Augmentation frame rates were chosen empirically and up to 8-fold data augmentation was evaluated. Test and validation set features were always extracted at the baseline frame rates.

\section{Results and Discussion}
\label{sec:results_disc}

The effectiveness of the proposed approach is demonstrated in three stages -- first, it is shown that training a model with FrAUG is better than the baseline approach of single frame rate training. Then, the performance of the proposed method is compared to conventional data augmentation techniques and lastly, the generalizabilty of the proposed method is evaluated by applying it to a different dataset with a different backend system and  different input acoustic features. Model performance is reported in terms  of the F1-score~\cite{chinchor1992muc} which is the harmonic mean of precision and recall. Statistical significance ($p<0.05$) was evaluated using the McNemar's test~\cite{mcnemar_note_1947}

\subsection{Multi Frame Rate Training}
\label{ssec:results1}

In the first set of experiments, DepAudioNet models were trained on the DAIC-WOZ dataset using different combinations of single frame rate and multiple frame rates. In this work, DepAudioNet was chosen as a baseline mainly because of DepAudioNet's open-source code~\cite{bailey2020raw}. The original paper where DepAudioNet was first proposed~\cite{ma2016depaudionet}  did not report test-set results because of the unavailability of ground truth labels and computed the F1-score for predictions at the speaker level and not at the frame level~\cite{rejaibi2019mfcc} nor the segment level~\cite{othmani2021towards}. Unfortunately, there is a lack of consensus regarding the evaluation protocols for the DAIC-WOZ dataset.

Performance comparison of single rate training versus multiple frame rate training on the validation set of the DAIC-WOZ dataset is shown in Table \ref{tab:results1_tab1}. The baseline frame rate of $L=64ms,R=50\%$ has an F1-score of 0.619. This is comparable to the reported F1-scores of 0.610 in prior works~\cite{ma2016depaudionet,bailey2020raw}. In contrast, the best performing configuration is the one with 5-fold data augmentation with multiple frame rate hyper-parameters of $L=64ms,128ms$ and $R=50\%, 25\%, 10\%$. Higher folds of data augmentation were also evaluated but 5-fold produced the best results. 

The best performing system has an F1-score of 0.656, a relative improvement of 5.97\% ($p = 4.72\mathrm{e}{-6}$) when compared to the baseline. This best performing configuration is better than any of the single frame rate performances, including when only the frame-widths are manipulated as in~\cite{padi2020multi}. A possible explanation for this result might be that a particular combination of time-frequency resolutions, provided to the model in FrAUG, contains depression-related information that is not available to the model when trained using single frame-rate features. 

\begin{table*}[hbt!]
\centering
\caption{Results, in terms of F1-score, comparing single frame rate training versus multi frame rate training using DepAudioNet and the DAIC-WOZ Validation set. $L$ and $R$ represent frame-width and frame-shift. $^*$ denotes the baseline F1-score. The best F1-score is boldfaced.}
\label{tab:results1_tab1}{%
\begin{tabular}{cccccccc}
\hline
\hline
$\downarrow$ L\textbackslash R $\rightarrow$ & 50\%  & 25\%  & 10\%  & 50\%, 25\% & 50\%, 10\% & 25\%, 10\% & 50\%, 25\%, 10\% \\ \hline \hline
32ms              & 0.601 & 0.604 & 0.569 & 0.606      & 0.633      & 0.562      & 0.604            \\
64ms              & 0.619$^*$  & 0.638 & 0.587 & 0.612      & 0.620      & 0.599      & 0.613            \\
128ms             & 0.648 & 0.627 & 0.588 & 0.618      & 0.628      & 0.638      & 0.616            \\
32ms, 64ms        & 0.637 & 0.607 & 0.579 & 0.615      & 0.617      & 0.623      & 0.602            \\
64ms, 128ms        & 0.635 & 0.612 & 0.576 & 0.620      & 0.625      & 0.617      & \textbf{0.656}   \\
32ms, 128ms        & 0.623 & 0.633 & 0.590 & 0.647      & 0.610      & 0.602      & 0.615            \\
32ms, 64ms, 128ms   & 0.626 & 0.607 & 0.546 & 0.655      & 0.600      & 0.582      & 0.596            \\ \hline \hline
\end{tabular}%
}
\end{table*}

To evaluate the performance on the test set, the best performing configuration on the validation data was selected and compared with the baseline. From Table \ref{tab:results1_tab1}, the best performing system is the one with $L=64ms, 128ms$ and $R= 50\%, 25\%, 10\%$.  The test-set results comparing the baseline and the best model are shown in Table \ref{tab:results1_tab2}. In this case, the proposed approach has an F1-score of $0.478$ compared to the baseline score of $0.382$ resulting in an improvement of $25.13\%$ ($p=5.66\mathrm{e}{-6}$). 

The authors of this paper acknowledge that higher baseline F1-scores for the test set have been reported in~\cite{othmani2021towards,rejaibi2019mfcc}. However, a comparison cannot be made because, unlike the approach in this paper, those systems either employed a different evaluation protocol such as segment-level predictions or re-partitioned the train-validation-test splits. More importantly, the goal of this paper is to show that FrAUG can provide significant gains over a baseline with no such augmentation. 

\begin{table}[hbt!]
\centering
\caption{Results, in terms of F1-score, for depression detection on DAIC-WOZ test data comparing baseline performance and the best configuration of FrAUG selected using validation data performance. The best F1-score is boldfaced.}
\label{tab:results1_tab2}{%
\begin{tabular}{ccc}
\hline \hline
L,R Configuration & Avg F1 Score & \begin{tabular}[c]{@{}c@{}}Data\\ Augmentation\end{tabular} \\ \hline \hline
\begin{tabular}[c]{@{}c@{}}Baseline\\ L=64ms, R=50\%\end{tabular}& 0.382 & None       \\ \hline
\begin{tabular}[c]{@{}c@{}}L=64ms, 128ms\\ R= 50\%, 25\%, 10\%\end{tabular} & \textbf{0.479} & 5x        \\ \hline \hline
\end{tabular}%
}
\end{table}


\subsection{FrAUG versus Conventional Data Augmentation Methods}
\label{ssec:results2}

To compare FrAUG with conventional data augmentation techniques, DepAudioNet models were trained using the DAIC-WOZ dataset with FrAUG, noise augmentation~\cite{snyder2018x}, VTLP-based augmentation~\cite{xu2021speech,ma2019nlpaug}, speed perturbation and pitch perturbation~\cite{ko2015audio}. The noise augmentation method was similar to the one used in Kaldi. The MUSAN library~\cite{snyder2015musan} was used to augment every utterance with randomly chosen foreground noise samples at a randomly chosen SNR of 0,5,10, or 15 dB~\cite{snyder2018x}.  The VTLP augmentation was based on the nlpaug library~\cite{ma2019nlpaug} and the method proposed in~\cite{jaitly2013vocal}. Speed and Pitch perturbation were based on~\cite{ko2015audio} and implemented using the Librosa library. For every augmentation method, up to 8-folds of data augmentation was applied and the best performing configuration, based on the validation set, was selected. The results comparing these augmentation methods are presented in Table \ref{tab:results2_tab1}.

\begin{table}[hbt!]
\centering
\caption{Results, in terms of F1-score, for depression detection on the DAIC-WOZ dataset comparing proposed method with conventional data augmentation techniques. The best F1-score is boldfaced.}
\label{tab:results2_tab1}
\begin{tabular}{cccc}
\hline \hline
\begin{tabular}[c]{@{}c@{}}Augmentation \\ Strategy\end{tabular}     & Validation & Test & \begin{tabular}[c]{@{}c@{}}Data\\ Augmentation\end{tabular}  \\ \hline \hline 
\begin{tabular}[c]{@{}c@{}}Baseline \end{tabular} & 0.619      & 0.382 & None \\ \hline 
\begin{tabular}[c]{@{}c@{}}Noise~\cite{snyder2015musan} \end{tabular} & 0.579 & 0.477 & 7x \\ \hline \vspace{1mm}
VTLP~\cite{jaitly2013vocal}  & 0.630 & 0.462 & 3x \\ \hline 
Speed Perturbation~\cite{ko2015audio} & 0.639 & 0.431 & 3x \\ \hline
Pitch Perturbation~\cite{ko2015audio} & 0.648 & 0.431 & 5x \\ \hline
FrAUG & \textbf{0.656} & \textbf{0.479} & 5x \\ \hline \hline
\end{tabular}
\end{table}

As seen in Table \ref{tab:results2_tab1}, FrAUG is the best performing augmentation strategy for both validation and test sets.  On the validation set, FrAUG outperforms  noise augmentation by 13.2\% ($p=3.43\mathrm{e}{-6}$), VTLP by 4.1\% ($p=4.92\mathrm{e}{-6}$), speed perturbation by 2.7\% ($p=1.53\mathrm{e}{-6}$) and pitch perturbation by 1.2\% ($p=3.71\mathrm{e}{-5}$). On the test set, the proposed approach is comparable to noise augmentation, is better than VTLP augmentation by 3.7\% ($p=4.95\mathrm{e}{-6}$), speed perturbation and pitch perturbation by 11.1\% ($p=4.88\mathrm{e}{-6}$ and $p=4.64\mathrm{e}{-6}$, respectively). One possible explanation for these results is that VTLP, speech perturbation and pitch perturbation, alters the spectral shape and therefore might be preserving less information about the depressive state of the speaker. In case of noise augmentation, a domain mis-match between training and validation data (noisy vs clean) may be the reason for degraded performance.  This shows that FrAUG can serve as an effective data augmentation strategy for depression detection without interfering with task-related acoustic information.

\subsection{Extension to CONVERGE Dataset}
\label{ssec:results3}

To show that the proposed approach is independent of the dataset, the model and the input acoustic feature, it was evaluated on the CONVERGE dataset using embeddings extracted from a pre-trained x-vector system. These embeddings were then used to train a CNN model to classify utterances as cases (depressed) or controls (healthy). 3x, 5x and 8x data augmentation was applied.  


\begin{table}[hbt!]
\centering
\caption{Results, in terms of F1-score, for depression detection on the CONVERGE dataset using x-vector embeddings with a CNN classifier as the backend, with and without FrAUG. The best F1-score is boldfaced.}
\label{tab:results3_tab1}
\begin{tabular}{cccc}
\hline \hline
L,R Configuration & Validation & Test & \begin{tabular}[c]{@{}c@{}}Data\\ Augmentation\end{tabular} \\ \hline \hline
\begin{tabular}[c]{@{}c@{}}Baseline\\ (L=64ms,R=50\%)\end{tabular}              & 0.676          & 0.654 & None         \\ \hline
\begin{tabular}[c]{@{}c@{}}L=32ms, 64ms\\ R=50\%, 25\%\end{tabular}             & 0.705          & 0.712 & 3x          \\ \hline
\begin{tabular}[c]{@{}c@{}}L=32ms, 64ms\\ R=50\%, 25\%, 10\%\end{tabular}       & 0.720          & 0.719 & 5x          \\ \hline
\begin{tabular}[c]{@{}c@{}}L=32ms, 64ms, 128ms\\ R=50\%, 25\%, 10\%\end{tabular} & \textbf{0.739} & \textbf{0.739} & 8x \\  \hline \hline
\end{tabular}
\end{table}

The effectiveness of FrAUG as applied to the CONVERGE dataset is evident from the results presented in Table \ref{tab:results3_tab1}. In comparison to the baseline F1-score of 0.676 (validation) and 0.654 (test), the best performing configuration (8-fold augmentation) has a performance of 0.739 and 0.739, respectively. This is an improvement of 9.32\% on the validation set ($p=5.93\mathrm{e}{-6}$) and 12.99\% on the test set ($p=5.88 \mathrm{e}{-6}$). The performance of the model improves consistently with increasing amounts of training data. Even though the downstream model was trained on x-vector embeddings and not on the acoustic features themselves, FrAUG improves the classification performance. This is a rather significant outcome because this shows that FrAUG can be beneficial in improving system performance even when applied to downstream tasks after the pre-training step. An important implication of this result is that FrAUG can be applied irrespective of the model training style - supervised pre-training, training from scratch, etc. 


\section{Conclusion and Future Work}
\label{sec:conclusion}

In this paper, a data augmentation method, called FrAUG, was proposed for depression detection from speech. Training data were augmented with new feature samples created by varying the frame-width as well as the frame-shift parameters during feature extraction. Thus, the proposed approach did not modify vocal tract or voice source related parameters and hence preserved acoustic information that may be important for MDD modeling purposes.  The proposed method of data augmentation performed better than a baseline system with no augmentation and four commonly used data augmentation methods. Lastly, the generalizability of the said method was demonstrated by improvements in classification performance on a different dataset with a different model and different input features. 

FrAUG improved the classification performance of DepAudioNet~\cite{ma2016depaudionet} trained using mel-spectrograms on the DAIC-WOZ dataset, and of a downstream network trained with x-vector embeddings generated from a pre-trained model~\cite{snyder2018x} using MFCCs on the CONVERGE dataset. It can therefore be suggested that the proposed method is independent of the dataset, the input acoustic features, the model and the model training style. 

Frame rate based data augmentation can be reliably used to increase the amount of training data and might prove to be useful in the development of large-scale MDD screening systems.  In the future, FrAUG will be applied to other features such as the Voice Quality features~\cite{park2016speaker}, and the fusion of FrAUG with other types of data augmentation techniques will be analyzed.  Further, FrAUG will be evaluated for other para-linguistic applications such as emotion recognition, detection of Alzheimer's dementia, etc. 

\section{Acknowledgements}
\label{sec:ack}
This work was funded by the National Institutes of Health under the award number R01MH122569- Combining Voice and Genetic Information to Detect Heterogeneity in Major Depressive Disorder.


\bibliographystyle{IEEEbib}
\bibliography{refs}

\end{document}